\documentclass{aa}  
\usepackage{graphicx}
\usepackage{txfonts}
\usepackage{newtxtext,newtxmath}
\usepackage[colorlinks=true, linkcolor=blue, citecolor=blue, filecolor=blue, urlcolor=blue]{hyperref}
\usepackage[capitalize]{cleveref}
\usepackage{orcidlink}      
\usepackage[normalem]{ulem}

\begin{document} 

   \title{Probing the millisecond pulsar origin of the $\gamma$-ray excess \\in the Galactic centre with LISA}
   \titlerunning{Probing the MSP origin of $\gamma$-ray excess in the Galactic centre with LISA}

   \author{Valeriya Korol\inst{1,2}$^{\orcidlink{0000-0002-6725-5935}}$\fnmsep \thanks{These authors contributed equally to this work.}
          \and
          Andrei Igoshev\inst{3}$^{\orcidlink{0000-0003-2145-1022}}$\fnmsep \thanks{These authors contributed equally to this work.}
          }

   \institute{Max-Planck-Institut für Astrophysik, Karl-Schwarzschild-Straße 1, 85748 Garching bei München, Germany
    \and
            SRON, Netherlands Institute for Space Research, Niels Bohrweg 4, 2333 CA Leiden, The Netherlands \\
            \email{v.korol@sron.nl}
         \and
             School of Mathematics, Statistics and Physics, Newcastle University, Newcastle upon Tyne, UK, NE1 7RU\\
             \email{andrei.igoshev@newcastle.ac.uk}
             }

   \date{Received: 16 September 2025 | Accepted: 8 November 2025}

  \abstract{
The gigaelectronvolt $\gamma$-ray excess observed towards the Galactic centre remains unexplained. While dark matter annihilation has long been considered a leading explanation, an alternative scenario involving a large population of millisecond pulsars remains viable. Testing this hypothesis with electromagnetic observations is difficult, as pulsar searches in the bulge are strongly affected by interstellar scattering, high sky temperature, and source confusion. We investigate whether gravitational-wave observations with the Laser Interferometer Space Antenna (LISA) could provide an independent probe of the millisecond pulsar binary population in the Galactic bulge in the future.

We constructed synthetic populations of detached millisecond pulsar–white dwarf binaries under two illustrative formation scenarios: an accreted scenario, in which systems are deposited by disrupted globular clusters, and an in situ scenario, in which binaries form through isolated binary evolution. In both cases, only $10^{-5}$–$10^{-4}$ of the underlying bulge population is detectable by LISA. Still, even a few detections would imply tens to hundreds of thousands of unseen systems. Accreted binaries are expected to have lower chirp masses ($\sim$0.4\,M$_\odot$), while in situ binaries produce more massive companions ($\sim$0.9\,M$_\odot$), though part of this contrast reflects our modelling assumptions.
LISA will measure binary frequencies with high precision, but chirp masses can only be determined for the most massive or highest-frequency systems. Thus, identifying millisecond-pulsar binaries among the far more numerous double white dwarfs will be challenging, as their gravitational-wave signals alone are indistinguishable. However, coordinated follow-up with the Square Kilometre Array of LISA-selected targets could directly test the millisecond-pulsar explanation of the $\gamma$-ray excess.
 }


    \maketitle
%

\section{Introduction}

More than a decade ago, an excess of $\gamma$-ray emission from the direction of the Galactic centre was discovered in data from the Fermi Gamma-ray Space Telescope \citep{Goodenough2009arXiv,HooperGoodenough2011PhLB,HooperLinden2011PhRvD}. Multiple studies independently confirmed an excess of $\gamma$-ray emission, relative to state-of-the-art background modelling, peaking around 1--3~GeV. This emission extends to at least $10^\circ$ from the centre and exhibits approximately spherical symmetry. 
The emitting region roughly overlaps with the Galactic bulge and has a spatial extent of $\approx 3$~kpc. Significant efforts have been made to test the robustness of the $\gamma$-ray background models (e.g. \citealt{Boyarsky2011PhLB, Gordon2013PhRvD, Calore2015JCAP, Zhou2015}) and the gigaelectronvolt excess has withstood these tests. The leading explanation for this excess, first proposed by \cite{Goodenough2009arXiv}, is that it originates from dark matter annihilation.

A few years prior to the discovery of the gigaelectronvolt excess, \citet{Baltz2007ApJ} realised that high-energy emission from radio pulsars can mimic the spectral properties of dark matter annihilation. Because pulsars produce high-energy emission with roughly similar spectra peaking around 2~GeV, \citet{HooperGoodenough2011PhLB} put forward the idea that a large population of unresolved (millisecond) pulsars could be the astrophysical source of the $\gamma$-ray excess. This hypothesis has received significant support over the years \citep{Abazajian2011JCAP,Mirabal2013MNRAS,Bartels2016PhRvL}.

Millisecond radio pulsars (MSPs; see \citealt{Lorimer2008LRR} for a review), also known as recycled pulsars, are neutron stars (NSs) characterised by short spin periods ($P_{\rm s} < 30$\,ms) and small spin period derivatives ($\dot{P}_{\rm s} < 10^{-19}$\,s\,s$^{-1}$). They are thought to acquire these properties through stable mass transfer in low-mass X-ray binaries (LMXBs). During this accretion phase, the NS gains angular momentum and is spun up to millisecond periods—a process known as the recycling mechanism \citep{Alpar1982Natur,Bhattacharya1991PhR}. The accreted material also buries the NS’s magnetic field, reducing it to values typically of $10^8$–$10^{10}$\,G. 
Once accretion ceases, the NS begins to lose rotational energy through magnetic dipole radiation, becoming a $\gamma$-ray source. The $\gamma$-ray luminosity is proportional to the spin-down luminosity ($\dot{E} \propto \dot{P}_{\rm s} / P_{\rm s}^3$), meaning that short spin periods result in strong $\gamma$-ray emission. Despite their relatively weak dipolar fields, MSPs are among the brightest $\gamma$-ray sources in the sky. As the MSP ages, its spin period gradually increases over gigayear timescales, leading to a slow decline in $\gamma$-ray luminosity.

Detecting MSPs in the Galactic centre through radio observations has proven to be challenging. The region is heavily affected by interstellar scattering and dispersion, which smear out the rapid pulsations of MSPs and reduce their detectability in standard radio surveys \citep[e.g.][]{1997ApJ...475..557C}. In addition, the high sky temperature and source confusion in the dense stellar environment further complicate searches. As a result, the true size of the MSP population in this region remains unknown. To date, only seven radio pulsars have been discovered in its vicinity \citep[e.g.][]{Johnston2006MNRAS,Deneva2009ApJ,Bates2011MNRAS,Eatough2013Natur,Wongphechauxsorn2024MNRAS}. \citet{Zhao2020ApJ} identified 110 compact radio sources, and \citet{Zhao2022ApJ} discovered 64 hypercompact sources in the same area, some of which may be radio pulsars. Multi-wavelength analyses have provided indirect constraints on the pulsar population. For instance, \citet{Wharton2012ApJ} estimated that the inner parsec (i.e. a region $\approx$300 times smaller than the $\gamma$-ray excess area) around the Galactic centre could host up to $10^3$ active pulsars. 
Over the years, several studies have attempted to estimate the size of the MSP population that could be responsible for the $\gamma$-ray excess \citep{Hooper2016JCAP,Ploeg2017JCAP,Dinsmore2022JCAP}. A recent study by \citet{Holst2025PhRvD} derived a $\gamma$-ray luminosity function using field MSPs and concluded that if $\gamma$-ray luminosity remains constant over time, the Galactic bulge could host as many as $3.6 \times 10^4$ MSPs. If instead the MSP $\gamma$-ray luminosity decays with time, the estimated population size rises to $2 \times 10^5$ MSPs.

The NSs and MSPs are also promising sources of continuous gravitational-wave emission. If a rapidly spinning NS has even a small amount of ellipticity—either imprinted at birth, caused by starquakes, accretion from a companion, or sustained by internal magnetic fields—it will be characterised by a non-zero mass quadrupole. This leads to the emission of gravitational waves at a frequency twice the star's rotational frequency, typically in the tens to thousands of hertz range, which falls within the sensitivity band of ground-based detectors \citep{1978ApJ...222..281F}. However, these continuous signals are intrinsically weak, and at the distance of the Galactic centre, they remain well below the detection threshold of current detectors \citep[e.g.][]{2022ApJ...932..133A,2023LRR....26....3R,2023ApJ...952...55S}. Next-generation detectors, such as the Einstein Telescope and Cosmic Explorer, are expected to achieve the sensitivity required to probe these sources in the Galactic centre \citep[e.g.][]{2023ApJ...952..123P, BartelProfumo2024}. 

For NSs and MSPs in binaries with orbital periods less than about two hours, the gravitational-wave signal lies in the millihertz band and is accessible to space-based detectors, such as the Laser Interferometer Space Antenna \citep[LISA;][]{LISARedBook}, even at distances comparable to the Galactic centre and beyond \citep{2019MNRAS.489.4513S, 2020MNRAS.492.3061L, 2021MNRAS.502.5576K, 2022ApJ...937..118W}. In this article, we investigate whether LISA can probe the MSP population in binary systems in the Galactic bulge and thus provide an independent test of their proposed contribution to the $\gamma$-ray excess. We argue that because gravitational-wave detection is subject to entirely different selection effects than electromagnetic searches for MSPs, LISA offers a promising new way to test the MSP origin of the $\gamma$-ray excess. To illustrate this potential, we present the first forecasts for MSP binaries in the Galactic bulge detectable by LISA.

This article is structured as follows. In Section~\ref{sec:review}, we briefly review the literature on the formation of MSP binaries and summarise their gravitational-wave detectability with LISA. In Section~\ref{sec:method}, we describe the assumptions and methodology behind our population synthesis of the bulge. In Section~\ref{sec:results}, we present our forecasts for different bulge formation scenarios and discuss the likelihood of detecting and characterising the MSP population with LISA. Finally, in Section~\ref{sec:discussion}, we discuss our findings in a broader context, and we summarise our conclusions in Section~\ref{sec:conclusions}.


\section{Galactic millisecond pulsar binary population} \label{sec:review}

The Australia Telescope National Facility (ATNF) pulsar catalogue v~2.6.1\footnote{\url{https://www.atnf.csiro.au/research/pulsar/psrcat}} \citep{Manchester2005AJ} contains information on 624 pulsars with spin periods shorter than $P \leq 30$\,ms, most of which are classified as MSPs. A significant fraction of these pulsars are located in globular clusters. For example, the most recent compilation by Paulo Freire\footnote{\url{https://www3.mpifr-bonn.mpg.de/staff/pfreire/GCpsr.html}} summarises properties of 345 pulsars in 45 clusters, indicating that at least 55\% of all known MSPs reside in globular clusters. Whether formed dynamically in globular clusters or through isolated binary evolution in the field, MSPs can end up in a variety of binary configurations. Below, we introduce those configurations that are mentioned in this study, drawing primarily from the classification and discussion presented in \citet{Tauris2023book}.

Many MSPs are thought to descend from LMXBs, in which a NS accretes material from a companion star and is spun up to millisecond spin periods. A large fraction of known MSPs are found in binaries with helium-core white dwarf (He~WD) companions. Other systems host more massive carbon–oxygen-core WDs (CO~WDs) or oxygen–neon–magnesium-core WDs (ONeMg~WDs). Some MSPs reside in tight ultra-compact X-ray binaries (UCXBs) with orbital periods as low as a several minutes, while others are found in eclipsing systems known as spiders. These are typically categorised as black widows, with extremely low-mass companions ($< 0.1$\,M$_\odot$), or redbacks, with more massive companions ($0.1$–$0.4$\,M$_\odot$). Most recently, \citet{Yang2025} discovered an MSP with a massive companion (up to $1.6$\,M$_\odot$), most likely a stripped helium star (He star).

\begin{figure}
\includegraphics[width=\columnwidth]{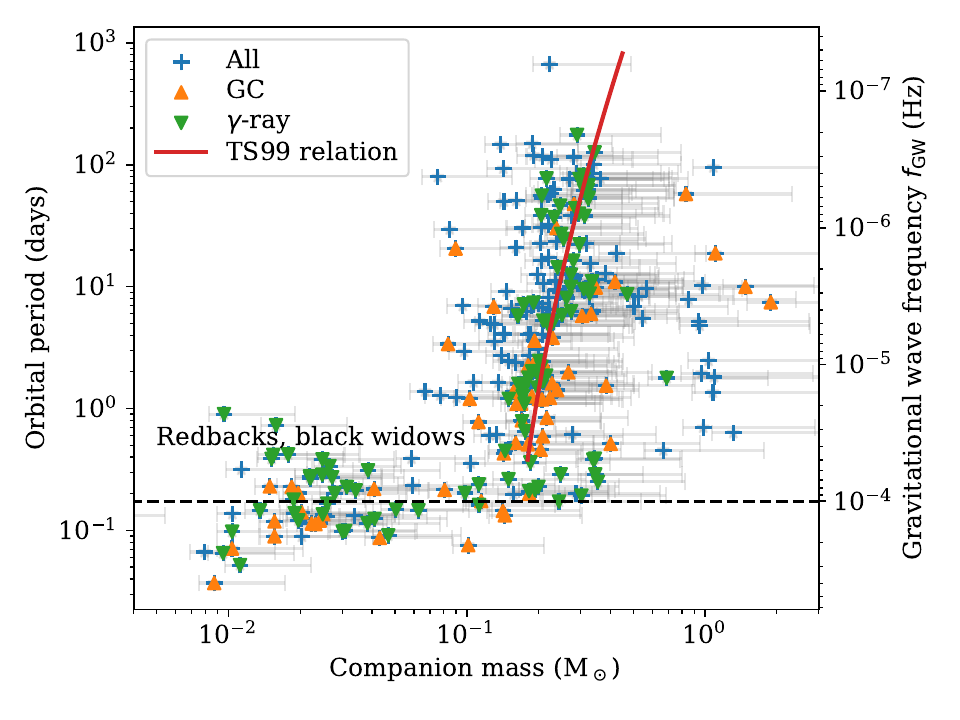}
\caption{Orbital periods and companion masses for MSPs compiled from the ATNF Pulsar Catalogue v2.6.0. The companion mass shown is the median mass, computed assuming an orbital inclination of $i = 60^\circ$. All data points (blue cross symbols) correspond to radio-detected MSPs. Among them, $\gamma$-ray sources are highlighted with green downward-pointing triangles, and those located in globular clusters (GCs) are marked with orange upward-pointing triangles. Grey error bars indicate lower and upper limits on WD companion masses. For reference, the theoretical orbital period–companion mass relation from \citet{TaurisSavonije1999AA} for Population I stars is shown as a solid red line. The minimum gravitational-wave frequency detectable by LISA is shown as a dashed black line. \\
}
\label{fig:orbital_periods}
\end{figure}

\begin{figure}
\includegraphics[width=\columnwidth]{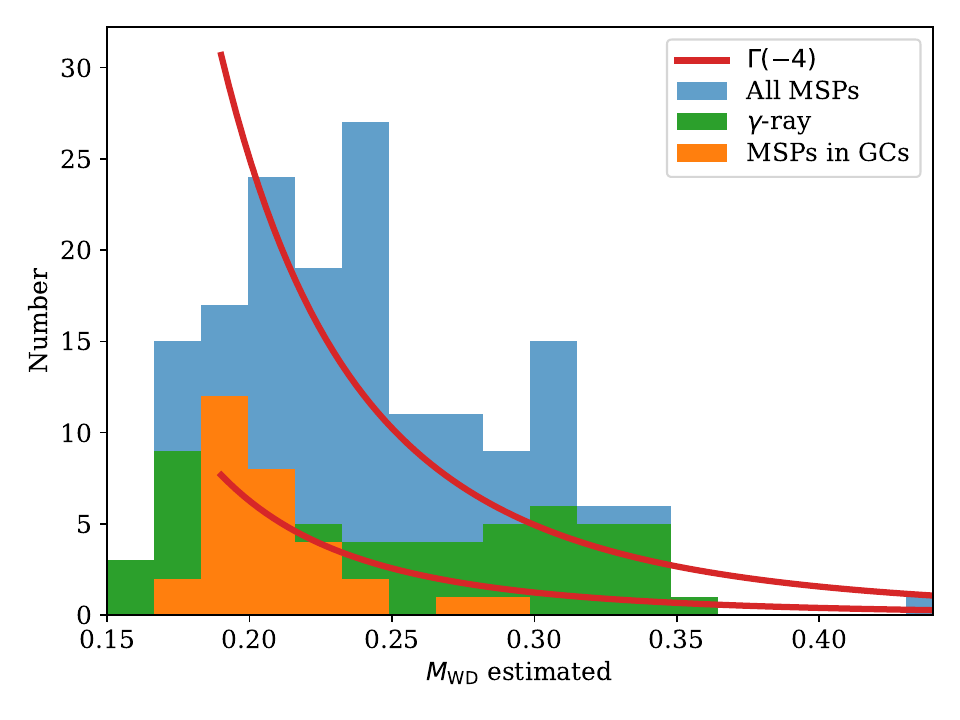}
 \includegraphics[width=\columnwidth]{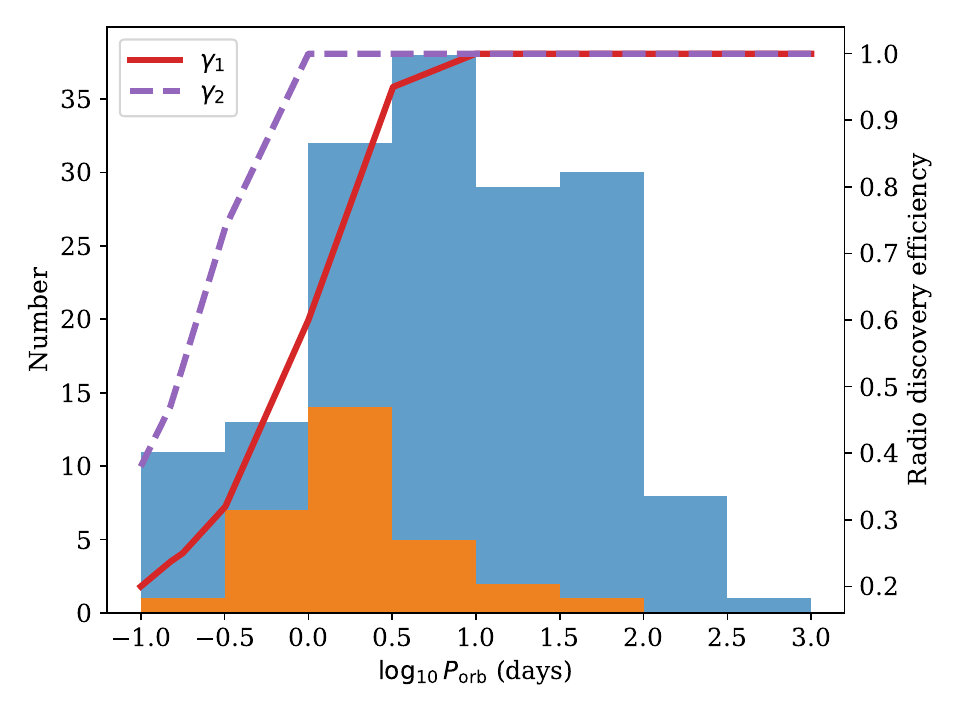}
\caption{
{\bf Top panel}: Distribution of companion masses for the MSP sample. The solid red lines show a power-law distribution with a slope of $-4$, plotted for comparison. 
{\bf Bottom panel}: Distribution of binary orbital periods for the MSP sample. Overplotted are the estimated pulsar detection efficiencies in radio surveys, based on the modelling by \citet{Bagchi2013MNRAS}. Two search strategies are shown: $\gamma_1$ for a standard (non-accelerated) search, and $\gamma_2$ for an acceleration search. These detection efficiency curves are computed assuming the following orbital parameters: eccentricity $e = 0$, $M_\mathrm{MSP} = 1.4$\,M$_\odot$, integration time $T\mathrm{obs} = 1000$\,s, WD companion mass derived from the \citet{TaurisSavonije1999AA} relation, spin period of 3\,ms, orbital inclination $i = 60^\circ$, and longitude of periastron $\Omega = 30^\circ$. 
}
\label{fig:orbital_periods_bottom_panels}
\end{figure}


\subsection{MSPs formation and orbital parameters}

In a comprehensive review, \citet{Tauris2011ASPC} summarised five basic scenarios for MSP formation. In all cases, mass transfer from a non-degenerate donor to a NS is essential. Two scenarios involve case A and B mass transfer in low-mass systems (donor mass $\lesssim 2$\,M$_\odot$), which lead to He~WD companions. The remaining three involve case A, B, and C mass transfer in intermediate-mass systems mass (donor mass $\sim$2–8\,M$_\odot$), resulting in CO~WD companions. In the latter, the shorter accretion phase results in only partially recycled pulsars with longer spin periods. 

\citet{TaurisSavonije1999AA} identified a relation between the mass of a He~WD and the orbital period at the end of stable mass transfer. This relation arises because stable mass transfer during the LMXB stage uniquely determines the orbital separation based on the donor star's density. The relation is given by
\begin{equation}
M_\mathrm{He WD} \; [M_\odot] = \left( \frac{P_\mathrm{orb}\; [\mathrm{days}]}{b} \right)^{1/a} + c,
\end{equation}
where $P_\mathrm{orb}$ is the binary orbital period.  For Population I stars, the coefficients are $a = 4.50$, $b = 1.2 \times 10^5$~days, and $c = 0.120$.
The mass of He~WDs typically lies between $0.15$ and $0.45$\,M$_\odot$. The slope of the relation flattens slightly at the lowest masses, as shown in a more recent work by \citet{Istrate2014AA}.

To better understand the observed MSP population, we selected all millisecond radio pulsars from the ATNF pulsar catalogue. We selected sources using the commonly adopted definition with $P_{\rm s} < 30$\,ms and $\dot{P}_{\rm s} < 10^{-19}$\,s/s. We also retrieved information about associations with globular clusters and the presence of the $\gamma$-ray emission. The resulting distribution of orbital periods and companion masses is shown in  Fig.~\ref{fig:orbital_periods}.
The MSPs fall broadly into two groups: (1) binaries with He~WD companions that follow the \citet{TaurisSavonije1999AA} relation (solid red line) and (2) binaries with low-mass companions—including systems that are the products of UCXB evolution, redbacks, and black widows.
The ATNF catalogue also includes some MSPs likely formed through a common-envelope channel with CO~WD companions, as well as rare systems in which MSPs are paired with NSs (i.e. those with companion masses $> 0.5$\,M$_\odot$ in Fig.~\ref{fig:orbital_periods}). Importantly, these latter systems are only partially recycled, meaning they have longer spin periods and higher magnetic fields. As a result, their spin-down luminosities are typically too low to produce detectable $\gamma$-ray emission, and these pulsars are therefore not seen as $\gamma$-ray sources. Nevertheless, some of them may be prominent gravitational-wave sources because of their larger component masses.
 
In Fig.~\ref{fig:orbital_periods_bottom_panels} we show the distributions of companion masses (top) and orbital periods (bottom) for the MSPs in our ATNF catalogue selection. The companion mass distribution can be approximated by a power law with an exponent of $-4$, which matches the observed trend down to $\sim 0.2$\,M$_\odot$, particularly for systems in globular clusters (orange distribution). In the right panel, we compare the orbital period distribution with the detection efficiency curves from \citet{Bagchi2013MNRAS} for standard ($\gamma_1$) and acceleration ($\gamma_2$) radio searches. The number of known MSPs drops sharply at short orbital periods, closely following the shape of the $\gamma_1$ efficiency curve. This suggests that observational biases significantly affect the detection of binaries with orbital periods shorter than one day, and implies that the intrinsic population may be more uniformly distributed in logarithmic orbital period space than the observed sample indicates.


\subsection{Detection with LISA}
\label{s:detection}

While traditional radio pulsar searches are limited in their ability to detect short-period binaries due to orbital acceleration effects (see bottom panel of Fig.~\ref{fig:orbital_periods_bottom_panels}), LISA will offer a complementary window into the MSP binary population. In particular, LISA will be sensitive to systems with orbital periods below one hour across the entire Milky Way \citep{2023LRR....26....2A}. In the case of circular binaries, as considered here due to the orbit being circularised during the recycling phase via stable mass transfer, the system emits continuous, quasi-monochromatic gravitational waves.
The gravitational-wave signal from such a binary can be described by eight parameters:
\begin{equation} 
\{ \mathcal{A}, f_{\rm GW}, \dot{f}_{\rm GW}, \lambda, \beta, \iota, \psi, \phi_0\},
\end{equation} \label{eqn:GWparams}
where $\mathcal{A}$ is the gravitational-wave amplitude, $f_{\rm GW} = 2/P_{\rm orb}$ is the gravitational-wave frequency (twice the orbital frequency), and $\dot{f}_{\rm GW}$ is its time derivative. The parameters $(\lambda, \beta)$ denote the ecliptic longitude and latitude, respectively; $\iota$ is the inclination angle of the orbital plane; $\psi$ is the polarisation angle; and $\phi_0$ is the initial phase of the waveform \citep[e.g.][]{2023arXiv231101300L}.
The gravitational-wave amplitude is given by
\begin{equation} 
\mathcal{A} = \frac{2 (G \mathcal{M})^{5/3} }{c^4 d} (\pi f_{\rm GW})^{2/3}, 
\end{equation} \label{eqn:amp}
where $G$ is the gravitational constant, $c$ is the speed of light, and $d$ is the distance to the binary. This shows that the gravitational-wave amplitude depends on the binary’s frequency, distance, and chirp mass ($\mathcal{M}$), defined as
\begin{equation} \mathcal{M} = \frac{(m_1 m_2)^{3/5}}{(m_1 + m_2)^{1/5}}, \end{equation}
where $m_1$ and $m_2$ are the binary component masses. 
The signal-to-noise ratio ($\rho$), averaged over time, ecliptic latitude, polarisation, and inclination, can be approximated as \citep[e.g.][]{2023MNRAS.522.5358F}:
\begin{equation} \label{eqn:snr}
\left\langle \rho^2 \right\rangle_{t,\psi,\beta,\iota} \approx \frac{24 \mathcal{A}^2 T_{\rm obs}}{25 S_n(f)}, 
\end{equation}
where $T_{\rm obs}$ is the observation time and $S_n(f)$ is the LISA noise power spectral density. In the following analysis, we assume the LISA noise power spectral density as given in the LISA Science Requirements Document \citep[][]{2021arXiv210801167B}. To assess a binary’s detectability, one can invert Eq.~\eqref{eqn:snr}, setting the observation time and a detection threshold. 

\begin{figure}
\includegraphics[width=\columnwidth]{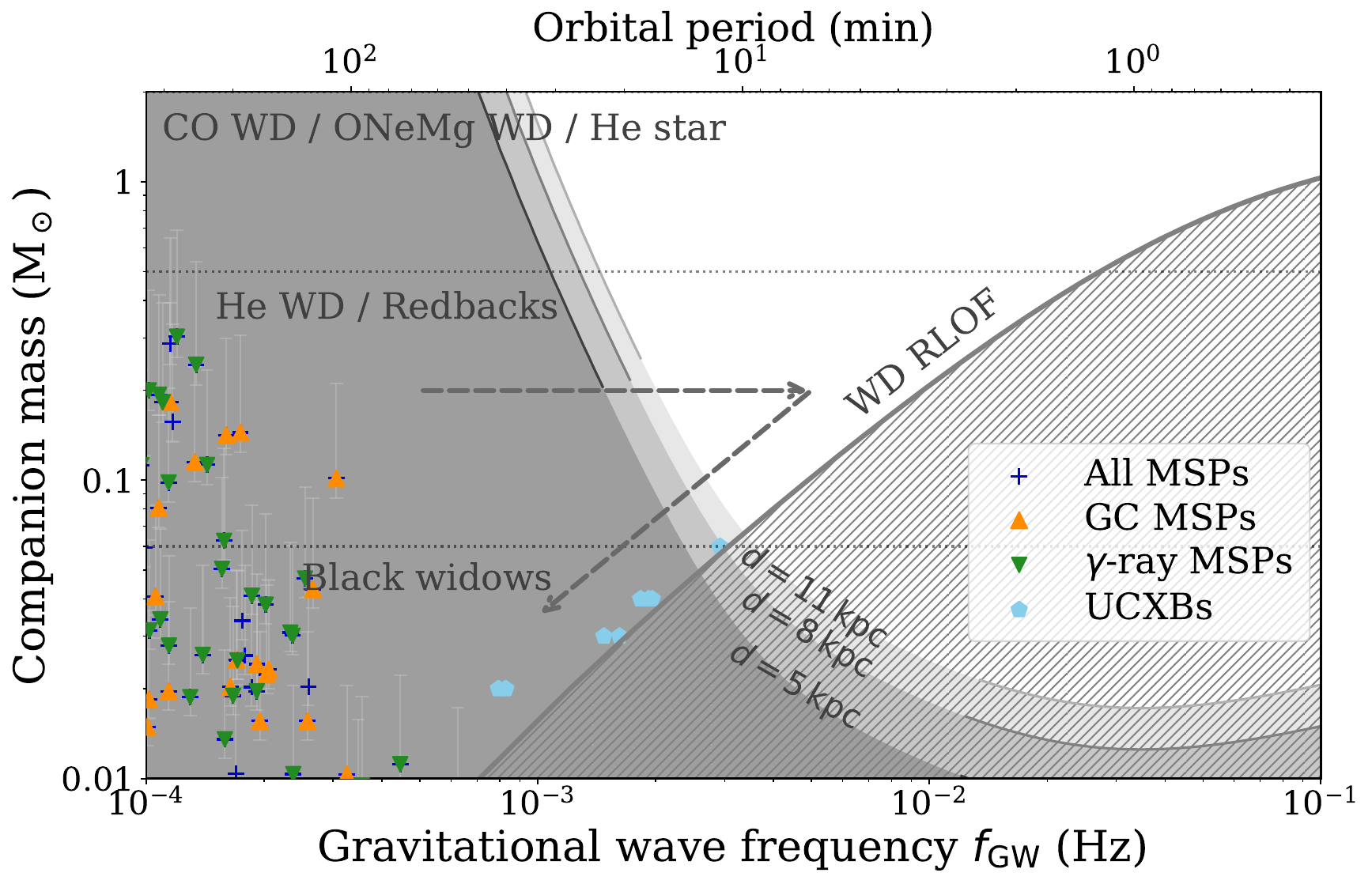}
\caption{Detectability of a binary system containing a 1.4\,M$_\odot$ MSP with LISA, shown as a function of companion mass and gravitational-wave frequency (bottom x-axis) or orbital period (top x-axis). Grey contours indicate LISA’s 4-year signal-to-noise threshold ($\rho_{\rm 4\,yr} = 7$) for sources located at 5 (dark grey), 8 (grey), and 11\,kpc (light grey). Dashed horizontal lines mark approximate boundaries between different classes of MSP binaries: black widows (low-mass companions, $M \lesssim 0.06$\,M$_\odot$), redbacks and He~WDs (intermediate-mass companions, $0.06 \lesssim M \lesssim 0.5$\,M$_\odot$), and more massive donors such as CO/ONeMg~WDs or He stars ($M \gtrsim 0.5$\,M$_\odot$). The grey solid curve marks the Roche-lobe overflow (RLOF) boundary for WD donors, indicating the onset of stable mass transfer. Observed MSP binaries are shown as coloured symbols, following the same classification as in Fig.~\ref{fig:orbital_periods}. UCXBs lying near the WD RLOF line are shown as light blue pentagons. The two-segment grey arrow illustrates the expected evolutionary track of close binaries. 
} \label{fig:detectability_plot}
\end{figure}

Figure~\ref{fig:detectability_plot} illustrates the detectable region in the companion mass–-gravitational-wave frequency parameter space, assuming an MSP mass of $1.4~\mathrm{M}_\odot$ and distances of 5 (dark grey), 8 (grey), and 11\,kpc (light grey). These distances span the inner Milky Way, which is relevant for the observed $\gamma$-ray excess \citep[e.g.][]{2013PDU.....2..118H}. The figure shows that LISA detectability at these distances requires frequencies of at least $\sim$1\,mHz (corresponding to orbital periods of less than $\sim$30\,min). For lower companion masses, the required frequency increases: approximately 3\,mHz for $\sim$0.1\,M$_\odot$, and up to $\sim$$10^{-2}$\,Hz for $\sim$0.02\,M$_\odot$.

While the detection of NS binaries in the inner 3\,kpc is, in principle, possible at gravitational-wave frequencies above a few millihertz, the companion star may—depending on its mass—fill its Roche lobe and initiate mass transfer before the system reaches the higher-frequency part of the LISA sensitivity window. As a reference, we plot in Fig.~\ref{fig:detectability_plot} the Roche lobe–filling radius of a WD donor using the commonly adopted zero-temperature radius–mass relation from \citet[][]{1988ApJ...332..193V}:
\begin{align}
    R_{\rm WD} =&\, 0.0114 \left[ 
        \left( M_{\rm WD}/ M_{\rm Ch} \right)^{-2/3} -  \left( M_{\rm WD}/ M_{\rm Ch} \right)^{2/3}  
        \right]^{1/2} \nonumber \\
        &\times   \left[  1 +
        3.5 \left( M_{\rm WD}/ M_{\rm p} \right)^{-2/3} -  \left( M_{\rm WD}/ M_{\rm p} \right)^{-1}  
        \right]^{-2/3},   
\end{align}
where $R_{\rm WD}$ is in R$_\odot$, $ M_{\rm Ch}=1.44\,$M$_\odot$ is the Chandrasekhar mass, and  $M_{\rm p}=5.7 \times 10^{-4}\,$M$_\odot$. This curve marks the maximum frequency at which detached WDNS binaries can exist, beyond which the donor will overfill its Roche lobe and initiate mass transfer. We also overplot the same subset of known MSPs as in Fig.~\ref{fig:orbital_periods}, as well as the shortest known UCXBs from \citet{2023A&A...677A.186A}.

Binary evolution simulations show that MSPs with He~WD companions initiate mass transfer at orbital periods of 5–7\,minutes \citep[e.g.][]{2017MNRAS.470L...6S}, after which the orbit begins to widen as the donor expands in response to mass loss. This sets an upper frequency limit for detecting such systems in the detached phase.
In contrast, simulations by \citet{2017MNRAS.467.3556B} show that systems with CO~WD or ONeMg~WD companions undergo unstable mass transfer when reaching these short orbital periods, leading to the tidal disruption of the donor.
Systems with He-star companions—i.e. stripped stellar cores still burning helium and thus larger and less compact than degenerate WDs—with masses up to approximately $1.2\,\mathrm{M}_\odot$ can initiate stable mass transfer at orbital periods between 10 and 80\,minutes, essentially set by the Roche lobe radius of the He~star \citep{2020ApJ...904...56G, 2021MNRAS.506.4654W}. The exact minimum period depends on the He star’s mass, with shorter periods corresponding to physically smaller (lower-mass) donors. Similar to systems with He~WD donors, these binaries evolve towards longer orbital periods after reaching the minimum. This characteristic evolution—initial inspiral driven by gravitational-wave emission, followed by mass transfer and orbital widening—is schematically illustrated with a grey two-segment arrow in Fig.~\ref{fig:detectability_plot}.
More massive He stars (above $1.2\,\mathrm{M}_\odot$) tend to initiate mass transfer that is too rapid to remain stable, often resulting in unstable mass transfer or mergers.


\section{Population synthesis of MSP binaries in the Milky Way bulge} \label{sec:method}

Many theories have been proposed to explain the origin of the Milky Way’s bulge \citep[e.g.][]{2017PASA...34...41N, 2016ASSL..418..233S}, each with distinct implications for the underlying stellar populations and binary formation channels.
Here, we consider two contrasting scenarios: one in which the bulge formed in situ as part of the Milky Way disc through internal evolutionary processes, and another in which it assembled through the accretion and disruption of globular clusters.
While these scenarios may be somewhat simplified, they provide a useful framework for exploring the potential properties and detectability of MSP binary populations in the bulge.

In the accretion scenario, the main idea is that, due to dynamical friction, globular clusters gradually spiral towards the Galactic centre. In the inner few kiloparsecs, the tidal field of the Galaxy can exceed a cluster’s self-gravity, stripping stars and potentially dissolving the cluster entirely \citep[e.g.][]{1975ApJ...196..407T, 1993ApJ...415..616C, 2012ApJ...750..111A, 2014MNRAS.444.3738A, Gnedin2014ApJ}. The globular clusters observed today are therefore likely the survivors of this process, representing only a small fraction of the original population. 
Globular clusters in the Milky Way halo are also known to be overabundant in MSPs (see Section~\ref{sec:review}). It is therefore reasonable to hypothesise that the MSPs responsible for the $\gamma$-ray excess could have originated in dense star clusters in the halo, which were subsequently disrupted, depositing their MSPs into the Galactic bulge. \citet{Brandt2015ApJ} showed that, in this scenario, the amplitude, angular distribution, and spectral characteristics of the resulting $\gamma$-ray emission are consistent with observations, suggesting that disrupted globular clusters could plausibly account for the observed excess in the Galactic centre.

In the in situ scenario, the Galactic bulge forms as part of the Milky Way disc through internal evolutionary processes \citep{2016ASSL..418..233S}. In this framework, early star formation is centrally concentrated, and the bulge emerges naturally from the dense, inner regions of the disc. Secular evolution may subsequently lead to the formation of a bar, which can undergo vertical buckling, producing the boxy or peanut-shaped morphology observed today \citep[e.g.][]{2015ASPC..491..169G, 2016PASA...33...27D, 2019A&A...629A..52L}.
This scenario is supported by a range of observational evidence, including a steep radial gradient in stellar ages—with stars in the inner few kiloparsecs typically older than those at larger radii—as well as metallicity and $\alpha$-element abundance patterns that suggest rapid early enrichment \citep[e.g.][]{2014A&A...562A..66Z, 2019MNRAS.490.4740B, 2020ApJ...900....4S}. These trends are consistent with a bulge dominated by an old, in situ stellar population that formed early in the Galaxy’s history and evolved alongside the disc.

In this section, we construct two model populations of MSP binaries in the bulge, each reflecting one of the two formation scenarios described above.


\subsection{Accreted scenario} \label{sec:accreted}

Here, we describe a simplified population synthesis model based on the assumption that the Galactic bulge was assembled through the accretion and disruption of multiple globular clusters. In this case, NS formation does not need to have occurred within the same binary system that later undergoes the LMXB phase. Instead, NSs may acquire new companions through dynamical interactions and subsequently evolve through the LMXB stage. For our purposes, we construct the population synthesis around the \citet{TaurisSavonije1999AA} relation, which links orbital period and companion mass for systems that have undergone stable mass transfer. This relation is relatively insensitive to the effects of dynamical evolution within clusters. Notably, MSPs observed in globular clusters follow this relation as well (see Fig.~\ref{fig:orbital_periods}).

In this scenario, a NS is born within a globular cluster and remains gravitationally bound to the cluster (see Section~\ref{sec:natal_kicks} for implications of natal kicks). At a later time, the NS acquires a low-mass companion ($M < 2\,$M$_\odot$) through dynamical interactions. The resulting binary undergoes stable mass transfer (i.e. an LMXB phase), during which the NS is spun up (i.e. recycled) to become an MSP, and the donor evolves into a He~WD. The system is expected to settle onto the \citet{TaurisSavonije1999AA} relation at the end of the mass transfer phase. Subsequently, the host globular cluster is accreted and merges with the central Galactic cluster, depositing the binary into the bulge. From this point onward, no further binary formation occurs. This scenario has several advantages. Binaries in which the NS does not undergo stable mass transfer retain long spin periods and exhibit low spin-down luminosities, making them unlikely contributors to the $\gamma$-ray excess observed in the Galactic centre. The \citet{TaurisSavonije1999AA} relation thus characterises the subset of He~WDNS binaries that have gone through the LMXB phase and are capable of producing observable MSPs.

Below we present the details of this population synthesis model.
Following the estimates of \citet{Holst2025PhRvD}, based on the observed $\gamma$-ray luminosity, we consider two benchmark values for the total number of MSPs (isolated and in binaries) in the Galactic bulge: $3.6 \times 10^4$ and $2 \times 10^5$. The lower value assumes that bulge MSPs have $\gamma$-ray luminosities similar to those observed in the field and in globular clusters, while the higher value corresponds to the case in which bulge MSPs are intrinsically fainter and therefore must be more numerous in order to produce the same emission.
To reflect the possibility that some recycled binaries may be disrupted in the dense environments of globular clusters after the LMXB phase, we assume a binary fraction of $0.5$, consistent with observations of MSP populations in both globular clusters and the field (see Section~\ref{sec:review}). Thus, we set $N^{\rm bulge}_{\rm MSP} = 1.8 \times 10^4$ and $1 \times 10^5$ for the number of MSPs in binaries in the bulge.

The initial mass of the NS is drawn from the bimodal normal distribution proposed by \citet{Antoniadis2016arXiv}, which reflects the observed mass distribution of MSPs. It is described as
\begin{align}
P(M_\mathrm{MSP})  = & \frac{1-r}{\sqrt{2\pi}\;\sigma_1} \exp\left[ -\frac{(M_\mathrm{MSP}-\mu_1)^2}{2\sigma_1^2} \right] \nonumber\\
& + \frac{r}{\sqrt{2\pi}\; \sigma_2}  \exp\left[ -\frac{(M_\mathrm{MSP}-\mu_2)^2}{2\sigma_2^2} \right].
\end{align}
The constants are as follows: $\mu_1 = 1.39$ M$_\odot$ and $\mu_2 = 1.81$ M$_\odot$, with standard deviations $\sigma_1 = 0.06$\,M$_\odot$ and $\sigma_2 = 0.177$\,M$_\odot$, and $r=0.4$.

To model the probability density function along the \citet{TaurisSavonije1999AA} relation, we explored two approaches. In the first, we drew the companion mass $M_\mathrm{WD}$ from a power-law distribution with slope $\gamma = -4$, denoted as $M_\mathrm{WD} \sim \Gamma(-4)$. As shown in the top panel of Fig.~\ref{fig:orbital_periods_bottom_panels}, this choice is consistent with observations. In the second approach, we drew the orbital period from a log-uniform distribution in the range 0.1–100 days, written as $P_\mathrm{\rm orb} \sim \log_{10} \mathcal{U}(-1, 2)$. Log-uniform distributions are commonly used in binary population synthesis and also appear plausible given current observational constraints (see bottom panel of Fig.~\ref{fig:orbital_periods_bottom_panels}).
In both cases, the model is expected to overproduce MSPs with the lowest WD masses and shortest orbital periods. This is an intended effect, as radio pulsar searches are known to be incomplete in the short-period regime. The sensitivity of radio searches degrades for MSP binaries with orbital periods below 1~day and He~WD mass below  0.19~M$_\odot$.  
Next, we use the \citet{TaurisSavonije1999AA} relation to estimate either the orbital period (if we draw $M_\mathrm{WD}$) or the WD mass (if we draw $P_\mathrm{orb}$). 

Next, we randomly paired each of the synthetic WDs with a NS, forming a population of WDNS binaries. This corresponds to the dynamical pairing of a low-mass star with a NS in the dense environment of a globular cluster, prior to the onset of stable mass transfer. None of these synthetic MSP systems would be immediately detectable by LISA, as their post-LMXB orbital periods remain relatively long ($P_{\rm orb}$ of approximately a few hours). However, we expect their orbits to gradually shrink over time due to the gravitational-wave radiation reaction.
To model this inspiral, we must assign an age to each system so that we can evolve its orbital parameters from the end of the LMXB phase to the present day. We adopt two approaches for estimating these ages. In the first, we draw ages from a normal distribution centred at 11\,Gyr with a standard deviation of 1\,Gyr, motivated by \citet{Jiaqi2025ARXIV}. In the second, we use the synthetic catalogue of \citet{ChenGnedin2024MNRAS}, randomly assigning to each system the time of accretion of an associated globular cluster.
We computed the orbital decay using the analytical prescription of \citet{Peters1964PhRv}, assuming zero eccentricity. This assumption is justified, as any significant eccentricity is expected to be damped during the LMXB phase due to stable mass transfer \citep{1992RSPTA.341...39P}.
We first converted the orbital period to a semi-major axis using Kepler’s third law:
\begin{equation}
a_0 = \left( \frac{G(M_\mathrm{WD} + M_\mathrm{NS})P_\mathrm{orb}^2}{4\pi^2} \right)^{1/3}.
\end{equation}
We then checked whether the binary merges within its assigned age by computing the merger time due to gravitational-wave emission:
\begin{equation} \label{eqn:tau_c}
T_c = \frac{a_0^4}{4\beta},
\end{equation}
where $\beta$ is given by
\begin{equation}
\beta = \frac{64}{5} \frac{G^3 M_\mathrm{WD} M_\mathrm{NS} (M_\mathrm{WD} + M_\mathrm{NS})}{c^5}.
\end{equation}
If the binary does not merge within its expected age, we computed the evolved semi-major axis as
\begin{equation} \label{eqn:a_f}
a_f = \left( a_0^4 - 4\beta t \right)^{1/4},
\end{equation}
and converted it back to an orbital period for use in our mock catalogue.

Finally, we assumed that binaries are spatially distributed according to a spherically symmetric Gaussian density profile, centred on the Galactic centre. The distribution is given by
\begin{equation} \label{eqn:bulge}
\rho_{\rm bulge}(r) \propto e^{-r^2 /r_{\rm b}^2}
\ \ \  \mathrm{kpc}^{-3},
\end{equation}
where $r$ is the spherical radial distance from the Galactic centre, and $r_{\rm b} = 0.5$\,kpc is the characteristic scale radius. 
This choice provides a simple approximation for the spatial extent of the bulge MSP population. It is sufficient for our purposes, as the detailed spatial structure of the bulge does not significantly affect detectability with LISA \citep{storck2023}.

\begin{table*}
    \centering
    \caption{
    Summary of the population synthesis models and predicted numbers of MSP binaries detectable with LISA.
    }
    \resizebox{\textwidth}{!}{%
    \begin{tabular}{lccccc}
    \hline \hline
    Name & Initial MSP distribution & Age distribution & $N_\mathrm{MSP}^\mathrm{bulge}$ & $N_\mathrm{MSP}^\mathrm{LISA}$ & $f^\mathrm{LISA}$ \\
    \hline
    Accreted A  & $M_\mathrm{WD} \sim \Gamma (-4)$ & $T \sim N (11, 1)$ & 100\,000 (18\,000) & 0 (0) & $<1 \times 10^{-5}$ \\
    Accreted B  & $P_\mathrm{orb} \sim \log_{10} \mathrm{U} (-1, 2)$ & $T \sim N (11, 1)$ & 100\,000 (18\,000) & 1 (0) & $1 \times 10^{-5}$ \\
    Accreted C  & $M_\mathrm{WD} \sim \Gamma (-4)$ & \cite{ChenGnedin2024MNRAS} & 100\,000 (18\,000) &  4 (1) & $4 \times 10^{-5}$ \\
    Accreted D  & $P_\mathrm{orb} \sim \log_{10} \mathrm{U} (-1, 2)$ & \cite{ChenGnedin2024MNRAS} & 100\,000 (18\,000) & 3 (1) & $3 \times 10^{-5}$ \\
    In situ $\gamma\alpha$  & \cite{Toonen2018AA}: $\alpha\lambda = 2,\,\gamma = 1.75$ & \cite{BP99} & $\sim$35\,000 & 4 & $1 \times 10^{-4}$ \\
    In situ $\alpha\alpha$  & \cite{Toonen2018AA}: $\alpha\lambda = 2$ & \cite{BP99} & $\sim$50\,000 & 4 & $8 \times 10^{-5}$ \\
    In situ $\alpha\alpha2$ & \cite{Toonen2018AA}: $\alpha\lambda = 0.25$ & \cite{BP99} & $\sim$4\,000 & 3 & $8 \times 10^{-4}$ \\
    \hline
    \end{tabular}
    }
    \tablefoot{ For each model, we list the initial MSP binary and age distributions, the total number of binaries in the bulge ($N_\mathrm{MSP}^\mathrm{bulge}$), the number of MSPWD systems detectable with LISA ($N_\mathrm{MSP}^\mathrm{LISA}$), i.e. those with $\rho_{4\mathrm{yr}}>7$, and the detection fraction $f^\mathrm{LISA}$, defined as the ratio of detectable systems to the total modelled bulge population. We use $\Gamma(p)$ to denote a power-law with exponent $p$, and $N(\mu, \sigma)$ for a normal distribution}
    \label{tab:models}
\end{table*}


\subsection{In situ scenario} \label{sec:in-situ}

The in situ formation scenario of WDNS binaries in the Milky Way was the focus of our previous study \citep{Korol2024MNRAS}. In that work, we constructed a present-day population of WDNS binaries by combining binary population synthesis models from \citet{Toonen2018AA}—which track stellar evolution from the main sequence to the WDNS stage—with a simple bulge+disc model of the Milky Way to assign spatial and age distributions. Based on the assigned ages, each binary was then evolved forward from the time of WDNS formation to the present day, as described in Section~\ref{sec:accreted} (see Eqs.~\eqref{eqn:tau_c}-\eqref{eqn:a_f}). 
Here, we briefly summarise the most relevant aspects of those models and focus on results for the bulge population. 

In this study, we consider three models from \citet{Korol2024MNRAS} that adopt different prescriptions for the common-envelope phase—the phase of rapid orbital shrinkage that occurs when one star expands to fill its Roche lobe and engulfs its companion.
Our model `in situ~$\alpha\alpha$' adopts the commonly used $\alpha$-formalism, which assumes that a fraction $\alpha$ of the orbital energy is used to eject the envelope, with the parameter $\lambda$ describing the envelope’s binding energy \citep{web84}. In this model, we assume a fixed value of $\alpha\lambda = 2$.
The model `in situ~$\alpha\alpha2$' also uses the $\alpha$-formalism, but with a lower efficiency parameter, $\alpha\lambda = 0.25$.
In contrast, the model `in situ~$\gamma\alpha$' adopts an alternative prescription for the common-envelope phase, known as the $\gamma$-formalism \citep{2000A&A...360.1011N}, which is based on the conservation of angular momentum rather than energy. In this model, the $\gamma$-prescription is applied unless the binary contains a compact object or the common-envelope phase is triggered by a tidal instability (rather than dynamically unstable Roche-lobe overflow). As a result, the first common-envelope episode is typically treated using the $\gamma$-formalism, while the second—typically involving a giant donor and a WD or NS companion—is modelled using the $\alpha$-formalism \citep[see][]{too12}. For this model, we adopt $\alpha\lambda = 2$ and $\gamma = 1.75$.
Our choice of efficiency parameter values is guided by previous studies that calibrated them using observations. The $\alpha\alpha$ and $\gamma\alpha$ models follow the prescriptions constrained by observed double WD systems \citep[hereafter WDWD; e.g.][]{2000A&A...360.1011N,nel05,vdS06}, while the lower efficiency in the $\alpha\alpha2$ model is based on calibration from compact WDs in binaries with M-type main-sequence companions \citep{zor10,too13,cam14,zor14}. 
In all three models, we adopt the NS natal kick prescription from \citet{Verbunt2017AA} as a distribution which describes observation of isolated radio pulsars well.

Our Milky Way model follows the approach of \citet{2004MNRAS.348L...7N}, with the star formation rate linearly interpolated from the plane-projected grid computed by \citet{BP99}, which is based on a spectro-photometric model of Galactic stellar populations. To represent the stellar bulge, we increase the star formation rate by a factor of two within the inner 3\,kpc compared to the original \citet{BP99} prescription, and distribute the resulting binary population according to Eq.~\eqref{eqn:bulge}. This yields a present-day integrated stellar mass of approximately $2.6 \times 10^{10}\,$M$_{\odot}$ within this region.

This simple prescription is motivated by the inside-out formation scenario of the Galaxy and captures the key feature of early star formation in the inner regions. As a result, the median age of binaries in our bulge model is around 8–10\,Gyr. While the model does not account for detailed Galactic dynamics—such as a rotating, bar-shaped bulge formed through secular evolution—it still provides a reasonable approximation of the stellar content in the inner Galaxy. In particular, including bar dynamics would mostly redistribute stars and binaries in phase space without significantly affecting the total stellar mass or the age distribution in the bulge.

\section{Results} \label{sec:results}

\begin{figure}
    \centering
    \includegraphics[width=0.95\columnwidth]{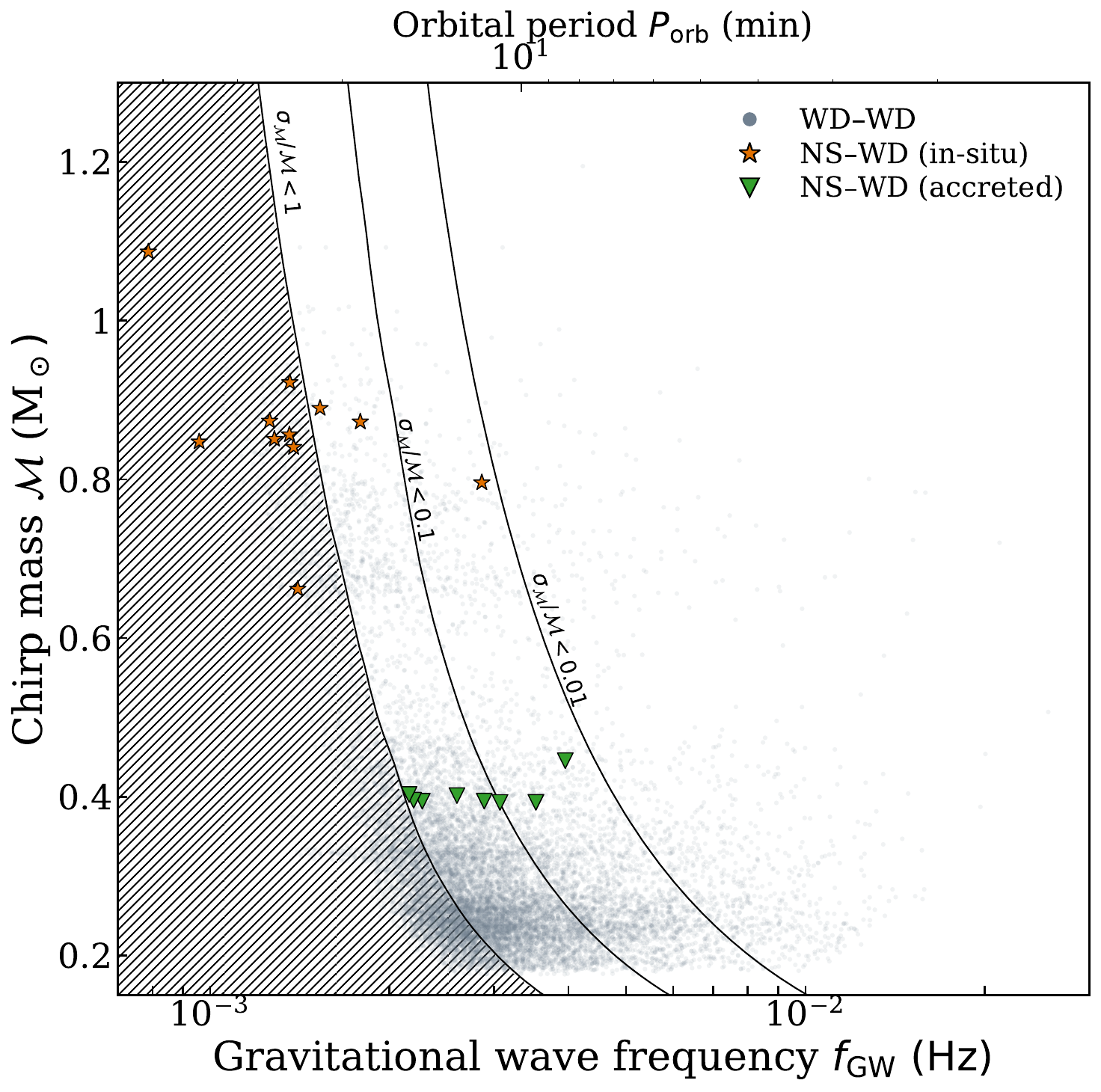}
    \caption{Detectable MSPWD binaries in the chirp mass–gravitational-wave frequency plane for both formation scenarios considered in this study. The in situ population is shown with orange stars, while systems from the accreted scenario are marked with green downward-pointing triangles. For comparison, the underlying population of detectable WDWD binaries is shown in blue-grey. Contours indicate regions of constant relative uncertainty in the chirp mass ($\sigma_{\mathcal{M}}/\mathcal{M}$) at levels of 1, 0.1, and 0.01, highlighting LISA’s ability to constrain binary parameters. The hatched region marks the part of parameter space where the chirp mass cannot be reliably measured ($\sigma_{\mathcal{M}}/\mathcal{M} > 1$). The top axis shows the corresponding orbital period.}
    \label{fig:chirp_mass_result}
\end{figure}

We present predictions for the number of MSP binaries in the Galactic bulge that could be detected by LISA, based on the population synthesis scenarios introduced in Section~\ref{sec:method}. Specifically, we explore two contrasting formation scenarios: one in which MSP binaries are deposited into the bulge through the accretion of disrupted globular clusters, and another in which they form in situ.
The key assumptions and results of each model are summarised in Table~\ref{tab:models}. For each scenario, we report 
the distributions used to set binary properties,
the total number of MSP binaries in the bulge ($N^{\rm bulge}_{\rm MSP}$), 
the number of binaries expected to be individually detectable by LISA ($N_{\rm MSP}^{\rm LISA}$),
and the corresponding detection fraction ( $f^{\rm LISA} = N_{\rm MSP}^{\rm LISA} / N_{\rm MSP}^{\rm bulge}$).
For the in situ models, the numbers of both the total bulge population and the detectable subset are estimated within the inner 1.7\,kpc, chosen to enclose 95\% of the bulge stars.
For accreted models, detectability is assessed using Eq.~\eqref{eqn:snr}, adopting a signal-to-noise ratio threshold of 7 for the nominal 4-year LISA mission duration. In this calculation, we also include the contribution of the Galactic gravitational-wave confusion foreground in the denominator, which is predominantly generated by the population of WDWD binaries. For the in situ models, the foreground has been estimated self-consistently in our previous work \citep{Korol2024MNRAS}, while for all accreted models we assume the foreground corresponding to the `in situ $\gamma\alpha$' model, which is also adopted in the LISA Definition Study Report \citep{LISARedBook}.

All models (except `Accreted A') yield at least one detectable system. The lack of detections in `Accreted A' arises from the smaller number of surviving systems at the present day, leaving no binaries above LISA’s detection threshold.
The fraction of detectable MSP binaries lies in the range $10^{-5}$–$10^{-4}$. Considering that LISA observations of Galactic binaries will be essentially complete at frequencies above a few millihertz \citep[e.g.][]{Korol2024}, even a single detection would indicate the presence of a much larger underlying population in the bulge—potentially numbering in the tens or hundreds of thousands. Conversely, a non-detection would begin to place upper limits on the size and properties of the bulge MSP binary population, helping to rule out the most optimistic models.
Since both the accreted and in situ scenarios are likely to contribute to the real population, the fact that their predicted detection fractions are of similar order suggests that LISA could detect binaries from either origin. Even a handful of detections would not only confirm the presence of a significant MSP binary population in the bulge, but also provide valuable insights into its formation history.

In Fig.~\ref{fig:chirp_mass_result}, we present the detectable MSP binaries in the chirp mass–-gravitational-wave frequency parameter space overplotting all models listed in Table~\ref{tab:models}. 
It is evident that the two scenarios populate distinct regions of this space, particularly in chirp mass. This difference primarily reflects the assumed or resulting WD masses in each model. In the accreted scenario, the MSP is paired by construction with a He WD, whose mass is drawn from the \citet{TaurisSavonije1999AA} relation. As a result, the chirp masses cluster around $\sim$0.4\,M$_\odot$. In contrast, in the in situ scenario, the companion mass emerges from the population synthesis. These binaries often contain more massive CO or ONe~WD companions, resulting in higher chirp masses, typically in the range $\sim$0.7–1.1\,M$_\odot$, because binaries with more massive companions are more likely to survive the common-envelope phase. Although the separation between in the chirp mass produced in the two scenarios may appear very clear, we caution that it is partly driven by the simplified assumptions adopted in each case. Specifically, the accreted models enforce the \citet{TaurisSavonije1999AA} relation, producing only low-mass He WD companions, while the in situ models likely underproduce such extremely low-mass WDs owing to relatively weak magnetic braking prescriptions \citep[e.g.][]{che21}.

We estimated the precision with which LISA can measure the parameters of MSPWD binaries using the Fisher matrix formalism \citep[e.g.][]{Cutler1998, TakahashiSeto2002, 2020MNRAS.492.3061L}. The gravitational-wave frequency is the most robustly measured parameter, with a relative uncertainty given by
\begin{equation} \label{eqn:sigmaF}
\frac{\sigma_{f_{\rm GW}}}{f_{\rm GW}} \simeq 8.7 \times 10^{-7} \left( \frac{f_{\rm GW}}{2\,{\rm mHz}} \right)^{-1} \left( \frac{\rho}{10} \right)^{-1} \left( \frac{T_{\rm obs}}{4\,{\rm yr}} \right)^{-1}.
\end{equation}
The chirp mass, $\mathcal{M}$, can be inferred when both the frequency, $f_{\rm GW}$, and its time derivative, $\dot{f}_{\rm GW}$, are both measurable. These two quantities are related by
\begin{equation} \label{eqn:fdot}
\dot{f}_{\rm GW} = \frac{96}{5} \pi^{8/3} \left( \frac{G\mathcal{M}}{c^3} \right)^{5/3} f_{\rm GW}^{11/3},
\end{equation}
which allows one to invert for $\mathcal{M}$. The relative error on $\dot{f}_{\rm GW}$ is given by
\begin{equation} \label{eqn:sigma_fdot}
\frac{\sigma_{\dot{f}_{\rm GW}}}{\dot{f}_{\rm GW}} \simeq 0.39 \left( \frac{f_{\rm GW}}{2\,{\rm mHz}} \right)^{-11/3} \left( \frac{\mathcal{M}}{0.9\,{\rm M}_\odot} \right)^{-5/3} \left( \frac{\rho}{10} \right)^{-1} \left( \frac{T_{\rm obs}}{4\,{\rm yr}} \right)^{-2}
\end{equation},
and propagating uncertainties yields
\begin{equation} \label{eqn:sigmaMe0}
\frac{\sigma_{\mathcal{M}}}{\mathcal{M}} \simeq \frac{11}{5} \frac{\sigma_{f_{\rm GW}}}{f_{\rm GW}} + \frac{3}{5} \frac{\sigma_{\dot{f}_{\rm GW}}}{\dot{f}_{\rm GW}}.
\end{equation}
The above expressions illustrate that while the frequency is typically measured with high precision, the chirp mass measurement is very sensitive to the binary’s mass and frequency, which determines whether $\dot{f}_{\rm GW}$ is detectable. In particular, only the most massive and/or high-frequency systems with sufficiently large $\dot{f}_{\rm GW}$ yield precise chirp mass estimates. 
This is illustrated in Fig.~\ref{fig:chirp_mass_result}, where we show contours of constant relative uncertainty in the chirp mass ($\sigma_{\mathcal{M}}/\mathcal{M}$) at levels of 1, 0.1, and 0.01. The region where the chirp mass is not measurable (i.e. $\sigma_{\mathcal{M}}/\mathcal{M}>1$) is hatched. 
Among the detectable MSP binaries, those from the accreted scenario tend to lie at higher frequencies, which compensates for their lower chirp masses and makes chirp mass measurements more likely. In contrast, most of the in situ binaries have higher chirp masses but lower frequencies, placing them in the regime where $\dot{f}_{\rm GW}$ is too small to be detected. As a result, only a few in situ systems fall in the region where $\mathcal{M}$ can be measured with meaningful precision.

Although we have shown that MSP binaries in the bulge can be detected by LISA and that a subset may yield measurable chirp masses, they are not easily distinguishable from the far more numerous population of WDWD binaries. To illustrate this potential confusion, Fig.~\ref{fig:chirp_mass_result} includes a representative sample of synthetic WDWD binaries from the bulge population model of \citet{Korol2024MNRAS}, shown as semi-transparent grey-blue circles. While the WDWD chirp mass distribution peaks between 0.2 and 0.3\,M$_\odot$, they remain abundant even in the regions occupied by MSP binaries.

The confusion with WDWD binaries has been previously investigated by \citet{2024MNRAS.531.2817M} and \citet{Korol2024MNRAS}, who showed that these two populations can overlap significantly in the full parameter space used to characterise quasi-monochromatic signals in the LISA band. One possible discriminant is orbital eccentricity, which may be retained in non-recycled NS binaries. However, MSPs formed through stable mass transfer are expected to have very low eccentricities ($\lesssim 10^{-3}$), well below LISA’s sensitivity threshold \citep{2024MNRAS.531.2817M}. As a result, they will be difficult to distinguish from WDWDs based on LISA data alone.


\section{Discussion} \label{sec:discussion}

In this study, we explored the prospects for detecting MSP binaries in the Galactic bulge with LISA. We considered two contrasting formation scenarios—accretion from disrupted globular clusters and in situ formation via standard binary evolution—and showed that both are expected to contribute individually resolvable sources. While these scenarios were chosen to bracket the range of plausible formation pathways for the bulge population, the true bulge population is likely to reflect a combination of both. Recent chemo-dynamical studies of the inner Milky Way suggest that a significant fraction of bulge stars formed early and rapidly, in burst-like episodes at high redshift \citep[e.g.][]{BelokurovKravtsov2022,BelokurovKravtsov2023}. Although our in situ models do not explicitly include a bursty star formation history, they are designed to capture early formation and may therefore already approximate key features of these more complex scenarios.

Our accreted model is intentionally data-driven and simple. It focuses exclusively on binaries consistent with having He WD companions, which constitute the majority ($\approx 70$\% in the mass range $0.1$~--~$0.5$~$M_\mathrm{WD}$) of the observed MSP companions (see Section~\ref{sec:review}). As a result, it likely underestimates the full detectable population from the accreted scenario. In particular, systems with more massive CO or ONe WD companions — though expected to be rarer — would exhibit higher chirp masses and thus be more readily detectable with LISA. Our detection estimates for the accreted scenario should therefore be interpreted as conservative lower limits. Future work could explore this in more detail using detailed $N$-body simulations of globular clusters. 
On the other hand, extremely low-mass WDs are underrepresented in our in situ population synthesis because the models of \citet{Toonen2018AA}, which underlie this scenario, adopt relatively weak magnetic braking prescriptions. As shown by \citet{Istrate2014AA} and \citet{che21}, stronger magnetic braking can produce more MSP binaries with extremely low-mass WD companion with short orbital periods. Consequently, the lack of detections around $\sim$0.4\,M$_\odot$ in the in situ scenario likely reflects this modelling limitation rather than an intrinsic absence of such systems.

There is another MSP formation scenario that we have not considered in this study, in which bulge MSPs originate from the accretion-induced collapse of ONe~WDs. In this channel, the WD collapses into a NS after accreting $\sim$0.1\,M$_\odot$ and surpassing the Chandrasekhar limit \citep[e.g.][]{2010MNRAS.402.1437H, 2013A&A...558A..39T}. This process is expected to produce negligible natal kicks, potentially enhancing NS retention in the bulge or within globular clusters. Recently, \citet{Gautam2022NatAs} proposed that the Galactic centre $\gamma$-ray excess can be naturally explained by a population of MSPs formed through this mechanism. However, MSPs formed via accretion-induced collapse are observationally difficult to distinguish from those produced via standard recycling, and theoretical uncertainties remain regarding the stability of mass accretion and the possibility of explosive outcomes before collapse.

\subsection{Role of spider pulsars}

Figure~\ref{fig:orbital_periods} shows that a noticeable fraction of $\gamma$-ray sources occupy the region associated with spider systems, that is, they belong either to the class of redbacks or black widows. These spider pulsars are typically identified through radio eclipses, which distinguish them from general MSPWD binaries.

In our population synthesis, we do not model whether the NS’s radio emission is eclipsed by its companion. In this respect, we represent redbacks as MSPs with He WD companions of masses below $0.4$\,M$_\odot$, while black widows are not explicitly modelled because their formation pathways remain uncertain. For example, \citet{Chen2013ApJ} suggested that the fate of a binary depends primarily on the irradiation efficiency and beaming geometry of the NS, implying that redbacks do not evolve into black widows. On the other hand, \citet{GinzburgQuataert2020MNRAS} proposed that these two populations are evolutionarily linked.

As discussed above, LISA is expected to detect some WDNS binaries that may later be identified and characterised in the radio band as redbacks. While black widows could contribute to the $\gamma$-ray excess from the Galactic centre, they remain undetectable by LISA (see Fig.~\ref{fig:detectability_plot}). Given that about 30\,\% of MSPs with detected $\gamma$-ray emission in the ATNF pulsar catalogue have median companion masses below 0.1\,M$_\odot$ (i.e. belonging to black widow category), our method may therefore underestimate the total number of WDNS-like systems in the Galactic bulge by up to $\sim$30\,\%.

\subsection{Neutron star natal kicks} \label{sec:natal_kicks}

NSs are known to receive high velocities at birth due to asymmetric supernova explosions \citep{Lyne1994Natur}. For a recent review see \cite{Popov2025arXiv250901430P}. These natal kicks often reach hundreds of km\,s$^{-1}$, well above the escape velocity of a typical globular cluster. Yet, the observed abundance of MSPs in globular clusters suggests that a non-negligible fraction of NSs are retained. \citet{Verbunt2017AA} and \citet{Igoshev2020MNRAS} analysed young radio pulsars with precise astrometric measurements and found that approximately $5 \pm 3$\% receive kicks low enough to be retained in globular clusters.

\citet{Igoshev2021MNRAS} further argued that analyses based solely on isolated pulsars are biased towards higher kick velocities, as NSs born with small kicks are more likely to remain in binaries and be detected as X-ray binaries or binary radio pulsars. This conclusion has been strengthened by recent studies of Be X-ray binaries, whose orbital eccentricities suggest a distinct low-kick component in the NS natal kick distribution \citep{Valli2025arXiv}. Specifically, this third component is consistent with a Maxwellian distribution with a standard deviation of just $\sim$5 km\,s$^{-1}$—low enough that virtually all such NSs would be retained in globular clusters.

In this study, we adopted a pragmatic approach: we assumed that a sufficient number of NSs are retained in globular clusters, as supported by radio and $\gamma$-ray observations. Whether these NSs are retained due to receiving weak natal kicks or because they were born through accretion-induced collapse is not critical for our population synthesis. Both formation channels are consistent with our modelling as long as the resulting MSPWD binaries obey the \citet{TaurisSavonije1999AA} relation.

\subsection{Ultra-compact X-ray binaries}

The UCXBs represent the evolutionary phase following detached NSs (or MSPs) in binary systems with WD companions, once gravitational-wave radiation shrinks the orbit sufficiently to trigger mass transfer \citep[e.g.][]{Nelemans2010, Tauris2018PhRvL}.
These systems can reach very short orbital periods (i.e. high gravitational-wave frequencies) and can be detected by LISA in the bulge (see Section~\ref{s:detection}). For example, the highest-frequency UCXB in Fig.~\ref{fig:detectability_plot}, 4U~1820-30, would be detectable by LISA in less than two years of observations, with its parameters well measured within the nominal four-year mission lifetime \citep{2023MNRAS.522.5358F}. While UCXBs typically experience accretion over longer timescales, their orbital periods increase over time, eventually moving out of LISA’s sensitivity band (see Fig.~\ref{fig:detectability_plot}). 

We estimate the number of UCXBs in our accreted scenario by checking whether a given binary would merge within 2\,Myr of its assigned age. We adopt this 2\,Myr threshold based on detailed simulations by \citet{Tauris2018PhRvL}, whose figure~2 shows that UCXBs spend approximately this duration with gravitational-wave frequencies $f_\mathrm{GW} > 3$\,mHz. If a binary merges within 2\,Myr, we assume the system is currently an UCXB. For binaries that do not meet this criterion, we compute their detectability by LISA as described in Section~\ref{s:detection}. We find that up to three and four UCXBs can be found in `Accreted C' and `Accreted D' models, respectively, if $N_\mathrm{MSP}^\mathrm{bulge} = 1\times 10^5$. For $N_{\rm MSP}^{\rm bulge} = 1.8\times10^4$, at most one UCXB is expected.

In our in situ models, systems that evolve into MSP binaries with low-mass He~WD companions—and eventually into UCXBs—are rare in the adopted binary population synthesis simulations \citep{Toonen2018AA}, and we do not predict any of them to be detectable by LISA in the bulge (see Fig.~\ref{fig:chirp_mass_result}). This is primarily because binaries with less massive companions have lower orbital energy and are therefore more likely to merge during the common envelope phase. However, a study by \citet{che21} shows that adopting stronger magnetic braking prescriptions can expand the parameter space for forming such binaries, potentially increasing the size of the LISA detectable UCXB population.

\subsection{Synergies with electromagnetic and future gravitational-wave observatories}

If LISA detects tens of WDNS-like binaries in the direction of the Galactic bulge, these sources will become prime targets for confirmation through radio observations.
The Square Kilometre Array (SKA) is one of the most promising facilities for such follow-up studies.
A recent white paper by \citet{Schoedel2024arXiv} estimates that SKA1-Mid will be able to detect millisecond pulsars at the distance of the Galactic centre with a 7$\sigma$ threshold in just 10\,hours of integration time. These systems would appear as point-like radio sources. Moreover, since the orbital period will be known from gravitational-wave observations, acceleration searches can be significantly optimised. In this context, all Galactic centre LISA sources should be observed in the radio with the SKA.

An additional synergy comes from $\gamma$-ray observations. The upcoming Cherenkov Telescope Array (CTA) will offer unprecedented sensitivity to high-energy $\gamma$-ray sources in the inner Galaxy. A recent study by \citet{PhysRevD.107.103001} shows that CTA will be capable of testing the pulsar interpretation of the Galactic centre $\gamma$-ray excess by resolving individual MSPs or constraining their cumulative contribution through statistical analyses. If LISA detects a population of compact binaries consistent with MSPs in the bulge, and their location and distribution align with the $\gamma$-ray excess morphology, this would provide compelling multi-messenger evidence in support of a pulsar origin. Joint analysis of LISA and CTA data may therefore help disentangle the nature of the excess and distinguish astrophysical sources from alternative explanations such as dark matter annihilation.

\citet{BartelProfumo2024} have already shown that the detection of isolated MSPs via gravitational-wave radiation will become tangible with third-generation ground-based detectors such as the Einstein Telescope \citep{ET}. Looking ahead to space-based gravitational-wave missions beyond LISA, several next-generation observatories have been proposed in response to European Space Agency’s call for future mission concepts. Among these, $\mu$-Ares \citep{Sesana:2019vho} and LISAmax \citep{Martens_2023} aim to improve sensitivity in the millihertz regime by at least an order of magnitude compared to LISA. Such a gain in sensitivity would significantly increase the number of detectable MSP binaries in the Galactic bulge and provide tighter constraints on their frequency evolution and sky localisation.


\section{Conclusions} \label{sec:conclusions}

In this study, we have explored whether gravitational-wave observations with LISA could contribute to answering the long-standing open question about the origin of the gigaelectronvolt $\gamma$-ray excess observed towards the Galactic bulge. Although the leading explanation in the literature points to dark matter annihilation, an alternative scenario involving a large population of MSPs has not been ruled out. The major challenge in testing the MSP hypothesis is that detecting pulsars in the inner Galaxy is notoriously difficult with electromagnetic observations due to scattering and source confusion. We therefore investigated whether the binary MSP population in the bulge could be probed through gravitational waves with LISA independently of these limitations.

We showed that at distances of 5–11\,kpc (corresponding to the Galactic bulge), MSP binaries can be individually detected by LISA if they emit gravitational waves at frequencies above a few millihertz. This corresponds to orbital periods shorter than about 20 minutes. This regime is particularly interesting because it complements the capabilities of radio surveys, which are most sensitive to longer-period binaries and experience a sharp decline in detection efficiency for orbital periods below approximately one day. LISA therefore provides access to a region of parameter space that is currently inaccessible to electromagnetic observations.

To model the MSP binary population in the bulge, we considered two illustrative formation scenarios: one in which MSP binaries are deposited into the bulge through the accretion of disrupted globular clusters and another in which they form in situ. The accreted scenario was constructed using a data-driven approach, with companion properties drawn from the well-established \citet{TaurisSavonije1999AA} relation and variations in the binary age distribution. In contrast, our in situ scenario is entirely theory driven, as it is based on binary population synthesis models that explore different channels of common-envelope evolution. While simplified, these scenarios span a plausible range of binary properties and enabled us to examine how each might appear in LISA observations. In reality, the bulge MSP population likely includes contributions from both channels.

We find that both scenarios predict similar detection fractions, with the number of binaries detectable by LISA amounting to approximately $10^{-5}$–$10^{-4}$ of the underlying MSP binary bulge population. Although this fraction is small, even a single detection would imply the existence of a much larger unseen population, potentially numbering in the tens or hundreds of thousands. In the accreted scenario, MSP binaries typically have chirp masses of $\sim$0.4\,M$_\odot$, while the in situ scenario produces systems with higher chirp masses of $\sim$0.9\,M$_\odot$, though part of this contrast reflects our modelling assumptions.

We also examined the extent to which LISA could measure the properties of these systems. While the gravitational-wave frequency is generally measured with excellent precision, determining the chirp mass requires detecting the frequency derivative, $\dot{f}_{\rm GW}$, which is only feasible for high-mass and/or high-frequency binaries. As illustrated in Fig.~\ref{fig:chirp_mass_result}, only a subset of detectable MSP binaries are expected to yield measurable chirp masses.

The key challenge, however, lies in distinguishing MSP binaries from the far more numerous population of WDWD binaries. We illustrated this issue by comparing a representative sample of WDWD binaries in Fig.~\ref{fig:chirp_mass_result}. Although WDWD chirp masses typically peak at lower values, they remain abundant even in the regions of parameter space occupied by both accreted and in situ MSP binaries. As a result, while a small number of bulge MSP binaries are expected to be individually resolvable by LISA, distinguishing them from WDWDs will be the main obstacle in using gravitational-wave observations to test the MSP origin of the $\gamma$-ray excess. A promising strategy may be to use LISA detections as a targeted list for follow-up with the SKA, which could confirm the presence of MSPs in these binaries and provide critical evidence of, or against, their contribution to the $\gamma$-ray excess.


\begin{acknowledgements}
We thank Silvia Toonen for sharing WDNS models. We also thank Martyna Chru\'sli\'nska, Mario Cadelano, Stephen Justham, and Vasily Belokurov for fruitful discussions and suggestions. A.I. thanks Anastasia Frantsuzova for answering many questions. A.I. thanks the Royal Society for the University Research Fellowship URF\textbackslash R1\textbackslash 241531. \\
\end{acknowledgements}

\bibliographystyle{aa} 
\bibliography{biblio}

\end{document}